\documentclass[review]{elsarticle}
\usepackage{lipsum}

\newcommand\blfootnote[1]{%
  \begingroup
  \renewcommand\thefootnote{}\footnote{#1}%
  \addtocounter{footnote}{-1}%
  \endgroup
}

\usepackage{lineno,hyperref}
\modulolinenumbers[5]

\journal{Journal of \LaTeX\ Templates}

\begin{document}

\begin{frontmatter}

\title{The future of stellar occultations by distant solar system bodies: perspectives from the Gaia astrometry and the deep sky surveys}

\author{J.~I.~B.~Camargo$^{1,2}$}
\cortext[mycorrespondingauthor]{Corresponding author}
\ead{camargo@linea.gov.br}
\author{J.~Desmars$^3$}
\author{F. Braga-Ribas$^{4,2}$}
\author{R. Vieira-Martins$^{1,2}$}
\author{M. Assafin$^{5,2}$}
\author{B. Sicardy$^3$}
\author{D. B\'erard$^3$}
\author{G. Benedetti-Rossi$^{1,2}$}
\address{$^1$Observat\'orio Nacional / MCTIC,  Rua General Jos\'e Cristino 77, 20921-400, Rio de Janeiro, Brazil}
\address{$^2$LIneA,  Rua General Jos\'e Cristino 77, 20921-400, Rio de Janeiro, Brazil}
\address{$^3$LESIA / Observatoire de Paris, CNRS UMR 8109, Universit\'e Pierre et Marie Curie, Universit\'e Paris-Diderot, 5 place Jules Janssen, F-92195 Meudon C\'edex, France}
\address{$^4$Federal University of Technology-Paran\'a (UTFPR / DAFIS), Rua Sete de Setembro, 3165, CEP 80230-901, Curitiba, PR, Brazil}
\address{$^5$Observat\'orio do Valongo / UFRJ, Ladeira do Pedro Ant\^onio 43, RJ 20080-090, Rio de Janeiro, Brazil}

\begin{abstract}
Distant objects in the solar system are crucial to better understand the history and evolution of its outskirts. The stellar occultation technique 
allows the determination of their sizes and shapes with kilometric accuracy, a detailed investigation of their immediate vicinities, as well as 
the detection of tenuous atmospheres. The prediction of such events is a key point in this study, and yet accurate enough predictions are available 
to a handful of objects only. In this work, we briefly discuss the dramatic impact that both the astrometry from the Gaia space mission and the 
deep sky surveys -- the Large Synoptic Survey Telescope in particular -- will have  on the prediction of stellar occultations and how they may 
influence the future of the study of distant small solar system bodies through this technique.
\blfootnote{\url{https://doi.org/10.1016/j.pss.2018.02.014}}
\blfootnote{\copyright 2018. This manuscript version is made available under the CC-BY-NC-ND 4.0 license \url{http://creativecommons.org/licenses/by-nc-nd/4.0/}}
\end{abstract}


\end{frontmatter}

\linenumbers

\section{Introduction}

Stellar occultation is a powerful technique that allows the determination of sizes and shapes of distant solar system objects with kilometric accuracy 
\citep{2010Natur.465..897E} \citep{2011Natur.478..493S} \citep{2012Natur.491..566O} \citep{2013ApJ...773...26B} \citep{2015MNRAS.451.2295G} 
\citep{2017Natur.550..219O} \citep{2014Natur.508...72B} (leading to albedos, densities), an investigation of their immediate vicinities 
\citep{2017Natur.550..219O} \citep{2014Natur.508...72B} \citep{2015A&A...576A..18O} (telling about the presence of rings, satellites, jets), 
and that may reveal tenuous - down to few nanobars - atmospheres \citep{2011Natur.478..493S} \citep{2012Natur.491..566O} \citep{2017Natur.550..219O} 
\citep{2009Icar..199..458W}. 

Accurate predictions\footnote{Where and when, on the Earth, an occultation event can be observed.} of occultation events are the very first step for 
the full success in the use of the technique. The relevance of this step is such that its improvement inevitably -- and positively -- affects the 
future of the study of distant small solar system bodies through stellar occultations.

Thanks to the astrometry from the Gaia space mission and the deep sky surveys, a huge advance in this study is closer than ever. More specifically, 
the first will provide over 1 billion stars with unprecedented (sub milli- to micro-arcsecond) astrometric accuracy (see, for instance, 
\citep{2005ESASP.576....5M} \citep{2016A&A...595A...1G}), while the latter, like the Large Synoptic Survey Telescope (LSST), will provide 
images from which short-term accurate ephemerides\footnote{Ephemerides whose uncertainties, for 1-3 years after the most recent observation 
used in the determination of these ephemerides, are smaller than the angular size of the respective occulting bodies as seen from the Earth.} 
of faint (down to $r\sim24.5$ in the case of the LSST \citep{2009arXiv0912.0201L}) solar system bodies can be determined. As a result, this will 
lead us to milli-arcsecond (mas) level - or better - predictions to tens of thousands of TNOs as explained in the next two sections.

\section{The power of Gaia}

\begin{figure}
\begin{center}
\includegraphics[width=1.0\textwidth]{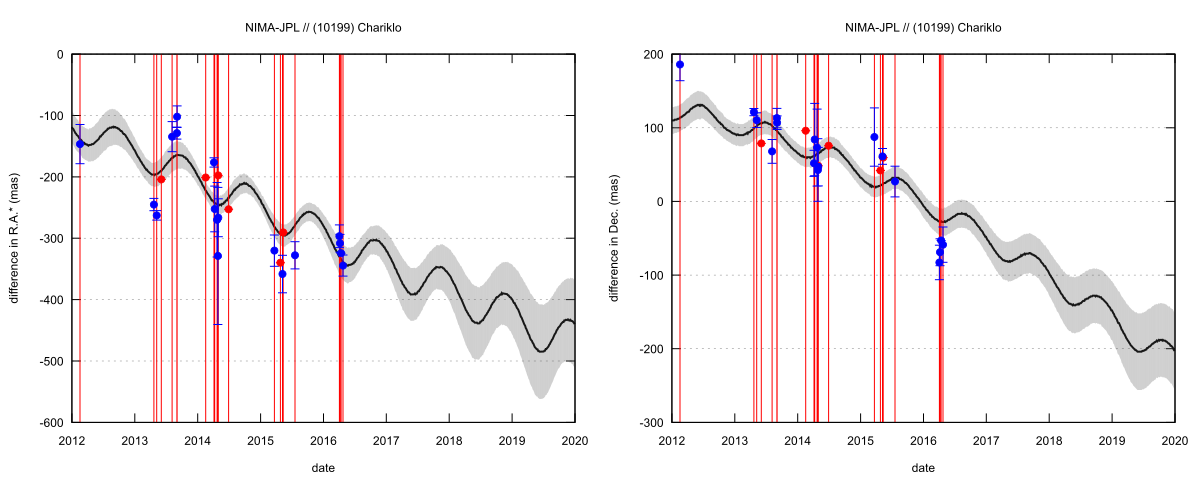}
\includegraphics[width=1.0\textwidth]{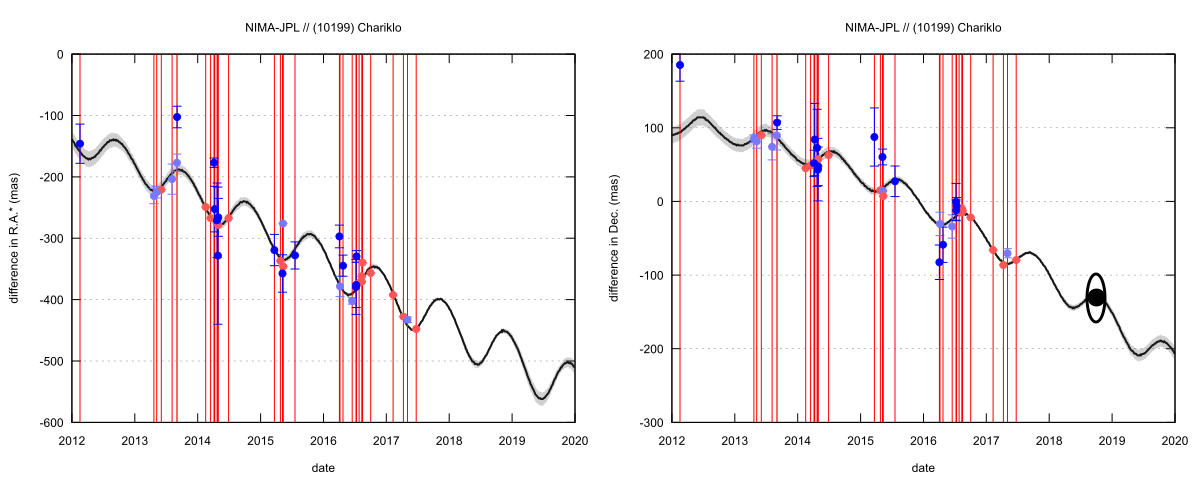}
\caption{Orbit improvement for Chariklo. Versions 8 (2016/JUN, upper panels) and 13 (2017/JUL, lower panels, the most recent one on the date 
this paper was written) of Chariklo's orbit as determined by NIMA. In all plots we have: differences in right ascension and declination 
(black curves) in the sense NIMA ephemeris minus the Jet Propulsion Laboratory (JPL) ephemeris (version: JPL20); 1-$\sigma$ uncertainty (grey area) 
of NIMA ephemeris; red points: differences between positions obtained from stellar occultations and those from the JPL ephemeris; dark/light blue 
points: differences between positions obtained from direct imaging of Chariklo and those from the JPL ephemeris; red vertical segments are 
provided for an easier visual correspondence between the dots and the dates associated to them; error bars represent the standard deviation
of the observational residuals from the same night and same observatory. Red dots, in particular, represent one observation each so that no 
error bar is attributed to them. In the upper panels, all positions are based on pre-Gaia astrometric catalogues. In the lower panels, red and 
light blue points  are now Gaia DR1-based. Some of these light blue points are a re-reduction of the non Gaia-based dark blue points in the upper 
panels. Note how well the Gaia-based positions agree with the orbit. Chariklo and its rings are also roughly represented in the lower right panel.}
\label{fig:ephem}
\end{center}
\end{figure}

 Figure~\ref{fig:ephem} shows the improvement in accuracy, thanks to Gaia, of the ephemeris of (10199) Chariklo as determined by our orbit fitting 
tool NIMA \citep{2015A&A...584A..96D}. In that figure, the most relevant difference between its upper and lower panels is that the first is based 
on pre-Gaia astrometry, whereas the latter is dominated by Gaia DR1 \citep{2016A&A...595A...2G} -based positions \citep{2016A&A...595A...4L} (but 
see also \citep{2017arXiv171010816K} for a foretaste of the impressive data that will be delivered by the Gaia Data Release 2 from April 2018).
 
 It should be noted that the uncertainty in the most recent version\footnote{On the date this paper was written.} of that ephemeris (lower panels) 
is significantly smaller than the angular size of Chariklo throughout 2018. And there are more accurate data and orbits to come! Note that all 
positions of Chariklo used here are from ground-based observations and that Gaia DR1 does not provide positions of small solar system bodies. 
Thousands of them, however, will be available in Data Release 2 \citep{2017arXiv171010816K},\citep{ACM2017Tanga}.
 
\begin{table}
\caption{Quantities from Fig.~\ref{fig:ephem}, lower panels}
\label{tab:quantities}
\begin{center}
\begin{tabular}{l r r r r r}     
\hline\hline
Technique & $\Delta\alpha{\rm cos}\delta$ & $\Delta\delta$ & Dots & Code & Gaia\\
\hline
Occultation        & 0  ($\pm$8)  & $-$1 ($\pm$4)  & 15 & red & Y\\
Direct imaging & 6 ($\pm$23)  & $-$3 ($\pm$10) &  8 & light blue & Y\\
Direct imaging  & 22 ($\pm$42) &   14 ($\pm$36) & 15 & dark blue & N\\
\hline\hline
\end{tabular}
\end{center}
Technique: way that positions were determined (from occultation or from direct imaging); $\Delta\alpha{\rm cos}\delta$ and $\Delta\delta$: average 
of the differences in the sense position minus NIMA. Values between parenthesis are the respective standard deviations; Dots: number of measurements 
used to derive the values in the two previous columns; Code: colour code as given in Fig.~\ref{fig:ephem}; Gaia: Gaia-based position? Yes/No. Angular 
measurements are in units of mas.
\end{table}
 
Currently, orbits determined by NIMA discriminates observational data (positions) of solar system bodies between those obtained from direct imaging 
and those obtained from occultations. It also discriminates between Gaia- and non Gaia-based positions. In the lower panels of Fig.~\ref{fig:ephem}, 
Gaia-based positions are given by the red and light blue points whereas non Gaia-based positions are given by the dark blue points. These discriminations 
are expressed in terms of weights.

\begin{figure}
\begin{center}
\includegraphics[width=0.8\textwidth]{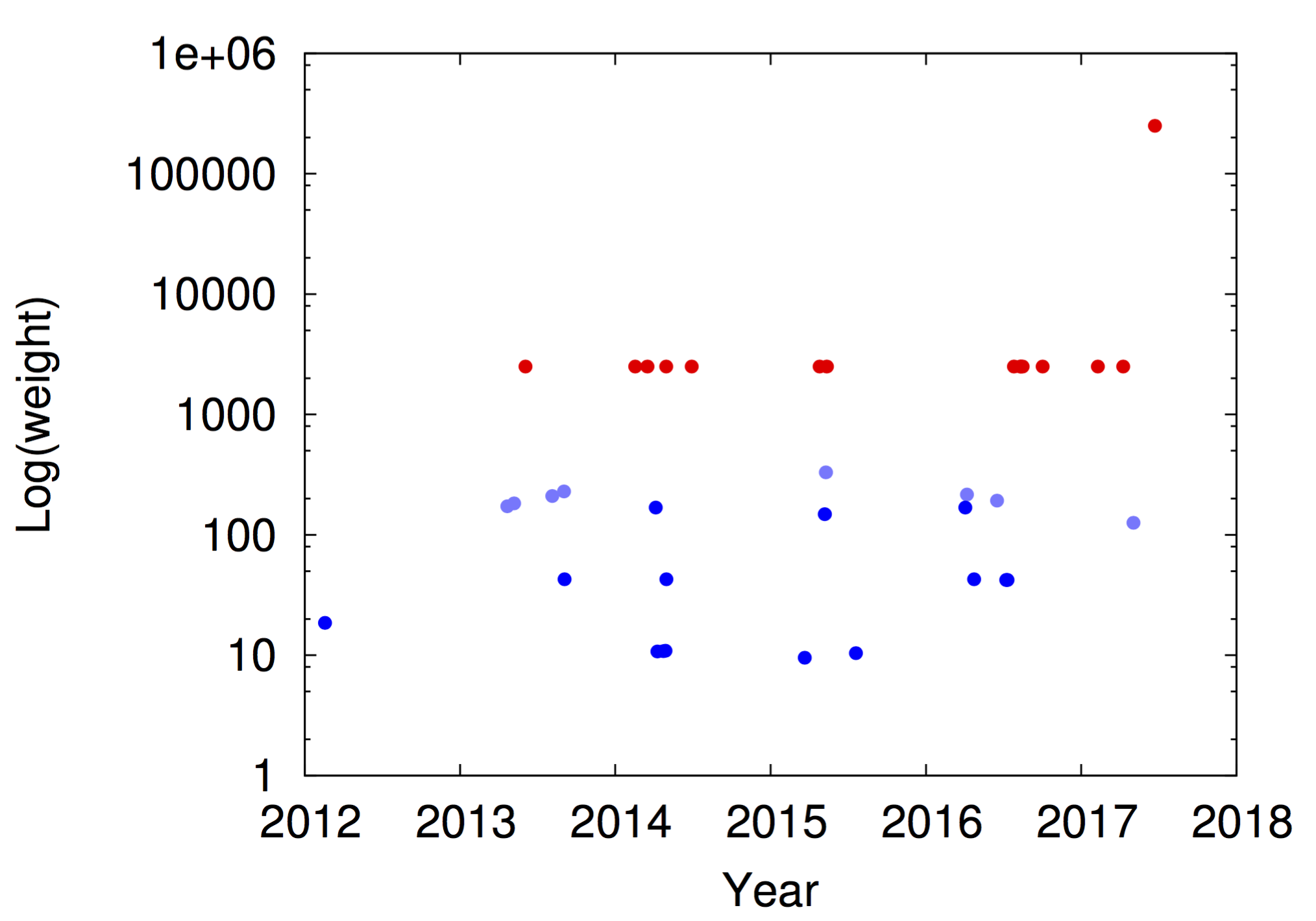}
\caption{Logarithm of the weights ($\sigma^{-2}$) attributed to the points shown in the lower panels of Fig.~\ref{fig:ephem}. The same colour code is used.}
\label{fig:weight}
\end{center}
\end{figure}

Figure~\ref{fig:weight} indicates the different weights attributed to the positions mentioned above. Astrometric data from an occultation is obtained 
from the relative position of the occulting body with respect to that of the occulted star. This relative position is determined with mas level accuracy 
(see, for instance, \citep{2011AJ....141...67S}). Therefore, accurate stellar positions, as those given by the Gaia mission, provide mas level accurate 
positions of the occulting (solar system) body. In this context, it is natural that these points (red dots in Fig.~\ref{fig:weight}) have the largest 
weights.

Table~\ref{tab:quantities} quantifies the effect of the weighing scheme (see \citep{2015A&A...584A..96D} for more details), that results in an orbit 
heavily dominated by the Gaia-based positions with emphasis to those from occultations.

\begin{figure}
\begin{center}
\includegraphics[width=1.0\textwidth]{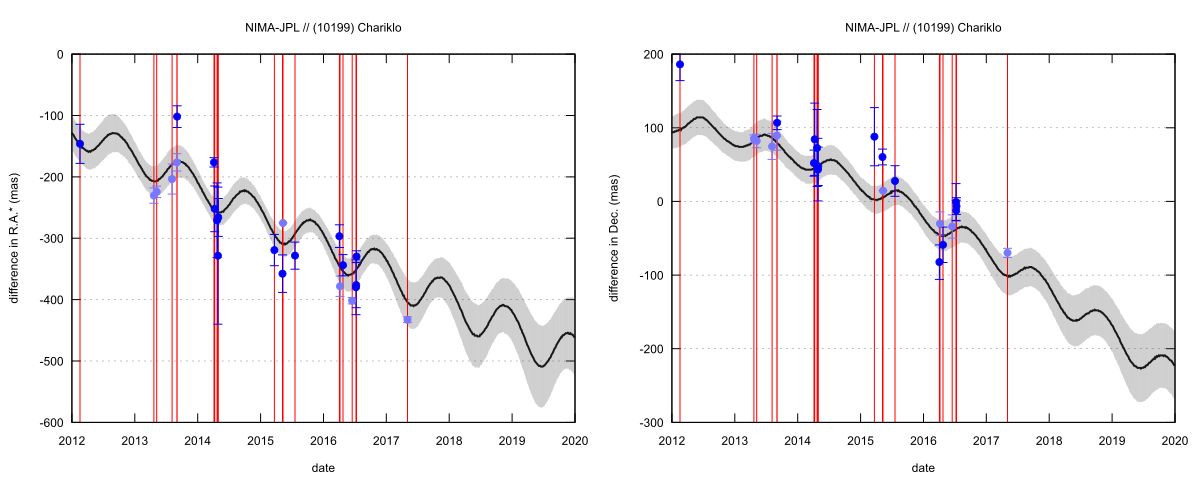}
\caption{Same as the lower panels of Fig.~\ref{fig:ephem}, but excluding the occultation data (red points) from the fit.}
\label{fig:noocc}
\end{center}
\end{figure}

The impact of the use of Gaia DR1-based positions from occultations can be seen in Fig.~\ref{fig:noocc}, from which the occultation data have been 
withdrawn. As compared to the lower panels of Fig.~\ref{fig:ephem}, we note not only a non negligible (at least in the context of predictions) difference 
between both ephemerides to those dates after that of the last observation but also a considerable increase in the uncertainty (grey zone). It is also 
important to note that, adding the occultation data, the ephemeris better fits the light blue (and also Gaia DR1-based) points, as expected. In this way, 
in the absence of occultation data, higher weights to the light blue points could be considered.

\section{The power of LSST}

The Minor Planet Center lists, to date, around 2\,600 transneptunian objects (TNOs) and Centaurs. The LSST, whose full science operations are scheduled to 
begin in 2023, will record the entire visible sky from Cerro Pach\'on about twice a week and observe millions of solar system objects, $\sim$40\,000 
TNOs among them \citep{2009arXiv0912.0201L}.

The survey expects to deliver astrometry accurate to 10 mas per exposure depending, among others, on the seeing and signal-to-noise ratio. In some cases, 
however, mas level accuracies may be required by stellar occultation predictions to observe satellites, grazing events by rings, topographic features, and 
bodies that may retain atmosphere. Therefore, a more careful astrometry may also be needed in specific cases and the previous section showed that 
this is possible.

It should be noted that many thousands of TNOs are expected to have more than one hundred observations along the ten years of operations of the 
survey \citep{2009arXiv0912.0201L}. These observations are crucial to, in association with the astrometry from Gaia, orbit fitting tools, and careful 
weighing schemes (e.g. \citep{2015A&A...584A..96D}), obtain accurate enough short-term ephemerides to all those objects.

\section{Comments and conclusions}

A stellar occultation event is magnitude independent, in the sense that it only needs to record the flux variation of the star in an interval of time that 
contains  its occultation. Therefore, the technique is suitable to also investigate the faintest occulting bodies. With Gaia and deep sky surveys we can expect 
the need to select events by focusing on, for instance, small groups of objects with different physical/dynamical features, instead of trying to observe 
them all. On the other hand, with the increasing amount of successful observations, we could also envisage a data-driven approach in the study of small 
solar system bodies uniquely from the occultations! The full profit of this exciting and quickly approaching scenario will, among others, rely on accurate 
predictions, on the amateur community, and on the availability of networks of small (20-40 cm) telescopes (e.g. \citep{2015EPSC...10..845B}) with fast readout 
detectors around the world, as well as the appropriate support to storage and data processing (big data context).

\section{Acknowledgments}

J.I.B.C. acknowledges CNPq grant 308150/2016-3. The work leading to these results has received funding from the National Institute of Science and Technology 
of the e-Universe project (INCT do e-Universo, CNPq grant 465376/2014-2). The work leading to these results has received funding from the European Research 
Council under the European Community's H2020 2014-2020 ERC grant Agreement n$^{\rm o}$ 669416 "Lucky Star". The authors acknowledge the comments from 
the anonymous referees.

\section{References}
\bibliography{mybibfile}

\begin{thebibliography}{10}
\expandafter\ifx\csname url\endcsname\relax
  \def\url#1{\texttt{#1}}\fi
\expandafter\ifx\csname urlprefix\endcsname\relax\def\urlprefix{URL }\fi
\expandafter\ifx\csname href\endcsname\relax
  \def\href#1#2{#2} \def\path#1{#1}\fi

\bibitem{2010Natur.465..897E}
J.~L. {Elliot}, M.~J. {Person}, C.~A. {Zuluaga}, et~al., {Size and albedo of
  Kuiper belt object 55636 from a stellar occultation}, Nature 465 (2010)
  897--900.
\newblock \href {http://dx.doi.org/10.1038/nature09109}
  {\path{doi:10.1038/nature09109}}.

\bibitem{2011Natur.478..493S}
B.~{Sicardy}, J.~L. {Ortiz}, M.~{Assafin}, et~al., {A Pluto-like radius and a
  high albedo for the dwarf planet Eris from an occultation}, Nature 478 (2011)
  493--496.
\newblock \href {http://dx.doi.org/10.1038/nature10550}
  {\path{doi:10.1038/nature10550}}.

\bibitem{2012Natur.491..566O}
J.~L. {Ortiz}, B.~{Sicardy}, F.~{Braga-Ribas}, et~al., {Albedo and atmospheric
  constraints of dwarf planet Makemake from a stellar occultation}, Nature 491
  (2012) 566--569.
\newblock \href {http://dx.doi.org/10.1038/nature11597}
  {\path{doi:10.1038/nature11597}}.

\bibitem{2013ApJ...773...26B}
F.~{Braga-Ribas}, B.~{Sicardy}, J.~L. {Ortiz}, et~al., {The Size, Shape,
  Albedo, Density, and Atmospheric Limit of Transneptunian Object (50000)
  Quaoar from Multi-chord Stellar Occultations}, ApJ 773 (2013) 26.
\newblock \href {http://dx.doi.org/10.1088/0004-637X/773/1/26}
  {\path{doi:10.1088/0004-637X/773/1/26}}.

\bibitem{2015MNRAS.451.2295G}
A.~R. {Gomes-J{\'u}nior}, B.~L. {Giacchini}, F.~{Braga-Ribas}, et~al., {Results
  of two multichord stellar occultations by dwarf planet (1) Ceres}, MNRAS 451
  (2015) 2295--2302.
\newblock \href {http://arxiv.org/abs/1504.04902} {\path{arXiv:1504.04902}}.

\bibitem{2017Natur.550..219O}
J.~L. {Ortiz}, P.~{Santos-Sanz}, B.~{Sicardy}, et~al., {The size, shape,
  density and ring of the dwarf planet Haumea from a stellar occultation},
  Nature 550 (2017) 219--223.
\newblock \href {http://dx.doi.org/10.1038/nature24051}
  {\path{doi:10.1038/nature24051}}.

\bibitem{2014Natur.508...72B}
F.~{Braga-Ribas}, B.~{Sicardy}, J.~L. {Ortiz}, et~al., {A ring system detected
  around the Centaur (10199) Chariklo}, Nature 508 (2014) 72--75.
\newblock \href {http://arxiv.org/abs/1409.7259} {\path{arXiv:1409.7259}},
  \href {http://dx.doi.org/10.1038/nature13155}
  {\path{doi:10.1038/nature13155}}.

\bibitem{2015A&A...576A..18O}
J.~L. {Ortiz}, R.~{Duffard}, N.~{Pinilla-Alonso}, et~al., {Possible ring
  material around centaur (2060) Chiron}, A\&A 576 (2015) A18.
\newblock \href {http://arxiv.org/abs/1501.05911} {\path{arXiv:1501.05911}},
  \href {http://dx.doi.org/10.1051/0004-6361/201424461}
  {\path{doi:10.1051/0004-6361/201424461}}.

\bibitem{2009Icar..199..458W}
T.~{Widemann}, B.~{Sicardy}, R.~{Dusser}, et~al., {Titania's radius and an
  upper limit on its atmosphere from the September 8, 2001 stellar
  occultation}, Icarus 199 (2009) 458--476.
\newblock \href {http://dx.doi.org/10.1016/j.icarus.2008.09.011}
  {\path{doi:10.1016/j.icarus.2008.09.011}}.

\bibitem{2005ESASP.576....5M}
F.~{Mignard}, {Overall Science Goals of the Gaia Mission}, in: C.~{Turon},
  K.~S. {O'Flaherty}, M.~A.~C. {Perryman} (Eds.), The Three-Dimensional
  Universe with Gaia, Vol. 576 of ESA Special Publication, 2005, p.~5.

\bibitem{2016A&A...595A...1G}
{Gaia Collaboration}, T.~{Prusti}, J.~H.~J. {de Bruijne}, A.~G.~A. {Brown},
  A.~{Vallenari}, C.~{Babusiaux}, C.~A.~L. {Bailer-Jones}, U.~{Bastian},
  M.~{Biermann}, D.~W. {Evans}, et~al., {The Gaia mission}, A\&A 595 (2016) A1.
\newblock \href {http://arxiv.org/abs/1609.04153} {\path{arXiv:1609.04153}},
  \href {http://dx.doi.org/10.1051/0004-6361/201629272}
  {\path{doi:10.1051/0004-6361/201629272}}.

\bibitem{2009arXiv0912.0201L}
{LSST Science Collaboration}, P.~A. {Abell}, J.~{Allison}, et~al., {LSST
  Science Book, Version 2.0}, (2009) ArXiv e-prints\href
  {http://arxiv.org/abs/0912.0201} {\path{arXiv:0912.0201}}.

\bibitem{2015A&A...584A..96D}
J.~{Desmars}, J.~I.~B. {Camargo}, F.~{Braga-Ribas}, et~al., A\&A 584 (2015)
  A96.
\newblock \href {http://dx.doi.org/10.1051/0004-6361/201526498}
  {\path{doi:10.1051/0004-6361/201526498}}.

\bibitem{2016A&A...595A...2G}
{Gaia Collaboration}, A.~G.~A. {Brown}, A.~{Vallenari}, et~al., A\&A 595 (2016)
  A2.
\newblock \href {http://dx.doi.org/10.1051/0004-6361/201629512}
  {\path{doi:10.1051/0004-6361/201629512}}.

\bibitem{2016A&A...595A...4L}
L.~{Lindegren}, U.~{Lammers}, U.~{Bastian}, et~al., {Gaia Data Release 1.
  Astrometry: one billion positions, two million proper motions and
  parallaxes}, A\&A 595 (2016) A4.
\newblock \href {http://arxiv.org/abs/1609.04303} {\path{arXiv:1609.04303}},
  \href {http://dx.doi.org/10.1051/0004-6361/201628714}
  {\path{doi:10.1051/0004-6361/201628714}}.

\bibitem{2017arXiv171010816K}
D.~{Katz}, A.~G.~A. {Brown}, {Gaia: on the road to DR2}, (2017) ArXiv
  e-prints\href {http://arxiv.org/abs/1710.10816} {\path{arXiv:1710.10816}}.

\bibitem{ACM2017Tanga}
P.~Tanga, {SMALL BODY OBSERVATIONS WITH GAIA}, ACM2017 - Abstract Book, (2017)
  http://acm2017.uy/abstracts/Plenary6.a.3.pdf.

\bibitem{2011AJ....141...67S}
B.~{Sicardy}, G.~{Bolt}, J.~{Broughton}, T.~{Dobosz}, D.~{Gault}, S.~{Kerr},
  F.~{B{\'e}nard}, E.~{Frappa}, J.~{Lecacheux}, A.~{Peyrot}, J.-P.
  {Teng-Chuen-Yu}, W.~{Beisker}, Y.~{Boissel}, D.~{Buckley}, F.~{Colas}, C.~{de
  Witt}, A.~{Doressoundiram}, F.~{Roques}, T.~{Widemann}, C.~{Gruhn},
  V.~{Batista}, J.~{Biggs}, S.~{Dieters}, J.~{Greenhill}, R.~{Groom},
  D.~{Herald}, B.~{Lade}, S.~{Mathers}, M.~{Assafin}, J.~I.~B. {Camargo},
  R.~{Vieira-Martins}, A.~H. {Andrei}, D.~N. {da Silva Neto}, F.~{Braga-Ribas},
  R.~{Behrend}, {Constraints on Charon's Orbital Elements from the Double
  Stellar Occultation of 2008 June 22}, AJ 141 (2011) 67.
\newblock \href {http://dx.doi.org/10.1088/0004-6256/141/2/67}
  {\path{doi:10.1088/0004-6256/141/2/67}}.

\bibitem{2015EPSC...10..845B}
M.~W. {Buie}, J.~M. {Keller}, L.~H. {Wasserman}, {RECON - A new system for
  probing the outer solar system with stellar occultations}, European Planetary
  Science Congress 10 (2015) EPSC2015--845.

\end{thebibliography}
\end{document}